\documentclass[10pt,conference]{IEEEtran}%
\usepackage{amsmath}
\usepackage{amsfonts}
\usepackage{amssymb}
\usepackage{graphicx}
\usepackage{amscd}%
\newtheorem{theorem}{Theorem}

\newtheorem{example}[theorem]{Example}

\begin{document}

\title{Testing goodness-of-fit via rate distortion}
\author{Peter Harremo\"{e}s\\Centrum voor Wiskunde en Informatica (CWI)\\Amsterdam, NL-1090 GB\\The Nederlands\\P.Harremoes@cwi.nl}
\maketitle

\begin{abstract}
A framework is developed using techniques from rate distortion theory in
statistical testing. The idea is first to do optimal compression according to
a certain distortion function and then use information divergence from the
compressed empirical distribution to the compressed null hypothesis as
statistic. Only very special cases have been studied in more detail, but they
indicate that the approach can be used under very general conditions.

\end{abstract}

\begin{keywords}
Bahadur efficiency, compact group, distortion, Gaussian distribution,
likelihood ratio test, rate-distortion function.
\end{keywords}

\section{Introduction}

There are many well-known examples of a fruitful interplay between information
theory and statistics. It started with \cite{Wald47} and \cite{Kullback51} and
is well described in \cite{CsisShi04}. Information divergence or Kullback
Leibler information plays a central role in measuring the distance between
probability distributions. Statistical testing is often delicate if the sample
size is small compared with size of the alphabet (sample space). If the
alphabet is a continuous set the normal approach in statistics is to
discretize the alphabet, but information is lost during discretization, and
often it is not clear how one should discretize the space.

Rate distortion theory was developed as a theoretical framework for lossy
compression. An obvious example is image compression, but rate distortion
theory often fails for this kind of application for three reasons. First of
all it is often very difficult to specify an appropriate distortion function.
Secondly, the statistics of the source is often not known. Thirdly, in most
cases it is impossible to calculate the rate distortion function exactly and
even a numerical calculation may be very involved due to the number of variables.

Although rate distortion theory was developed for lossy compression we claim
that the ideas are very useful for statistical analysis.

\section{Likelihood ratio testing}

On a finite sample space of size $k$ one can use information divergence as
statistics for testing goodness of fit. This is the likelihood ratio test. We
want to test a null hypothesis $H_{0}:P=P_{0}.$ An iid sample $\omega$ from
$P$ of size $n$ is made and the null hypothesis is accepted if $D\left(
Emp_{n}\left(  \omega\right)  \Vert P_{0}\right)  $ is smaller than some value
and rejected if it exceeds this value. The critical value is determined by the
significance level. If $P_{0}$ is the uniform distribution then $H\left(
Emp_{n}\left(  \omega\right)  \right)  =\log k-D\left(  Emp_{n}\left(
\omega\right)  \Vert P_{0}\right)  $ so in this case it makes no difference
whether one uses entropy or information divergence. Using large deviation
theory one will see that no other test is more Bahadur efficient than the
likelihood ratio test. The distribution of $2nD\left(  Emp_{n}\left(
\omega\right)  \Vert P_{0}\right)  $ will will converge to a $\chi^{2}$
distribution with $k$ degrees of freedom, so determining the values correspond
to different significance levels is simple.

This method cannot be used directly if the sample space is infinite and
$P_{0}$ is continuous. If $P_{0}$ is a distribution on $\mathbb{R}$ with
continuous distribution function $F$ then a popular method for testing
goodness of fit is the divide $\mathbb{R}$ into $k$ bins of equal probability.
As one want to keep points together if they are close on the real axis the
bins should be chosen of the form $\left[  F^{-1}\left(  \frac{j-1}{k}\right)
;F^{-1}\left(  \frac{j}{k}\right)  \right[  .$ If $f$ maps a point into its
bin then $P_{0}$ is mapped into a uniform distribution so we can use the
entropy $H\left(  f\left(  Emp_{n}\left(  \omega\right)  \right)  \right)  $
as statistic to test goodness of fit. The idea is then to increase the number
of bins slowly as $n$ increases. Recently it was proved that entropy is more
Bahadur efficient than other power statistics if $k$ is increased so slowly
that the mean number of samples per bin $n/k\ $tends to infinity for
$n\rightarrow\infty,$ see \cite{Harremoes2008, Harremoes2008e} and references
in there. This condition will hold if for instance $k=n^{1/2}$ and this choice
of number of bins will also ensure that distribution of entropy will be
asymptotically Gaussian.

It is easy to divide $\mathbb{R}$ into $k$ bins of equal probability for a
continuous distribution but it is not obvious how to do the same for
distributions on $\mathbb{R}^{2}$ or in higher dimensions. Even in one
dimension it is far from obvious why the bins should be of equal probability.
Maybe a different choice of bins would sometimes give a test that in one or
another sense is more efficient. To get better founded criteria for how to
choose bins we need a distortion function.

\section{The rate distortion test}

Consider a distribution $Q$ on a set $\Omega$ with a distortion function
$d:\Omega\times\Omega\rightarrow\mathbb{R}.$ For a distortion level $d_{0}$
the optimal coupling at distortion level $d_{0}$ is given by a Markov kernel
$\Psi_{d_{0}}:\Omega\rightarrow M_{+}^{1}\left(  \Omega\right)  .$ We shall
use $\Psi_{d_{0}}$ to smooth the empirical distribution so that we can compare
it with the null hypothesis $H_{0},$ i.e. we shall use $D\left(  \Psi_{d_{0}%
}\left(  Emp_{n}\left(  \omega\right)  \right)  \Vert\Psi_{d_{0}}\left(
Q\right)  \right)  $ as statistic for testing goodness of fit. There are
various ways to approximate $D\left(  \Psi_{d_{0}}\left(  Emp_{n}\left(
\omega\right)  \right)  \Vert\Psi_{d_{0}}\left(  Q\right)  \right)  $
numerically. We shall not discuss this problem. In general the rate distortion
function and $\Psi_{d_{0}}$ cannot be calculated exactly but using iterative
methods like the Arimoto Blahut algorithm they can be approximated. We shall
discuss three examples where the rate distortion function and $\Psi_{d_{0}}$
are given by explicit formulas.

\begin{example}
[Test of uniformity]We consider a set $A$ with $l$ elements. The set has no
particular structure so we use Hamming distortion as distortion function. Our
null hypothesis is $P=U$ where $u$ denotes the uniform distribution on $A.$ In
this case the Markov kernel $\Psi_{d_{0}}$ has the form
\[
\Psi_{d_{0}}:x\rightarrow\alpha\delta_{x}+\left(  1-\alpha\right)  U
\]
for some value $\alpha\in\left[  0;1\right]  $ determined by $d_{0}.$ The
Markov kernel maps the uniform distribution into the uniform distribution.
Therefore the statistic of the rate distortion test has the form%
\[
D\left(  \alpha Emp_{n}\left(  \omega\right)  +\left(  1-\alpha\right)  U\Vert
U\right)  .
\]
This statistic is closely related to the idea of \emph{local alternatives}
often studied in statistics.
\end{example}

\begin{example}
[Normality test]We consider the real numbers with squared Euclidian distance
as distortion function. Our null hypothesis is $P=\Phi$ where $\Phi$ denotes
the standard Gaussian distribution. The optimal Markov kernel for the rate
distortion problem sends $x$ into the distribution of $\alpha x+\left(
1-\alpha^{2}\right)  ^{1/2}Z$ where $Z$ is a standard Gaussian random
variable. We see that the Gaussian distribution is mapped into it self. Thus
the statistic of the rate distortion test is%
\[
D\left(  \alpha X+\left(  1-\alpha^{2}\right)  ^{1/2}Z\Vert\Phi\right)
\]
where we have identified the random variable
\[
\alpha X+\left(  1-\alpha^{2}\right)  ^{1/2}Z
\]
with its distribution. This Markov kernel can be rewritten as%
\begin{multline*}
D\left(  \alpha X+\left(  1-\alpha^{2}\right)  ^{1/2}Z\Vert\Phi\right) \\
=D\left(  X+\left(  \frac{1}{\alpha^{2}}-1\right)  ^{1/2}Z\Vert\Phi\left(
0,\alpha^{2}\right)  \right)
\end{multline*}
so the Markov kernels essentially smooth data by adding an independent
Gaussian random variable with variance $\alpha^{-2}-1.$ The idea of smoothing
data is well-known in statistics.
\end{example}

\begin{example}
[Test of uniformity of angular data]In this example we consider data with
values on the circle $s_{1}$ that we can identify with $\mathbb{R}%
/2\pi\mathbb{Z}.$ See \cite{Batschelet1981} for references. As distortion
function we shall use $4\cos^{2}\left(  \frac{\theta_{2}-\theta_{1}}%
{2}\right)  ,$ i.e. squared Euclidean distance between points on a circle. We
shall test the hypothesis $P=U$ where $U$ denotes the uniform distribution on
the circle. The optimal Markov kernel is a smoothing by adding a von Mises
distribution
\[
\frac{\exp\left(  \kappa\cos\left(  \theta\right)  \right)  }{2\pi
I_{0}\left(  \kappa\right)  }%
\]
where $I_{0}$ is the modified Bessel function of order $0$ with parameter
$\kappa$ determined by the distortion level \cite{Harremoes2006,
Harremoes2009b}. The Markov kernel maps the uniform distribution into the
uniform distribution.
\end{example}

\section{Limits for extreme values of $\beta$}

Often the rate distortion curve is parametrized by its slope $\beta.$ Here we
shall discuss the effect of choosing very small or very large values of
$\beta$ when the sample is kept fixed. We shall go through our three main
examples from this point of view.

\begin{example}
[Test of uniformity continued]Small or large values of $\beta$ corresponds to
small or large values of $\alpha.$ For $\alpha=1$ we get the statistic%
\[
D\left(  Emp_{n}\left(  \omega\right)  \Vert U\right)
\]
which is the likelihood ratio test. For $\alpha$ close to $0$ we use that
information divergence is an $f$-divergence with $f\left(  x\right)  =x\ln x$
so that%
\begin{multline*}
\frac{d}{d\alpha}D\left(  \alpha Emp_{n}\left(  \omega\right)  +\left(
1-\alpha\right)  U\Vert U\right) \\
=\frac{d}{d\alpha}\sum_{i=1}^{l}\frac{1}{l}f\left(  \frac{\alpha\hat{p}\left(
i\right)  +\left(  1-\alpha\right)  \frac{1}{l}}{\frac{1}{l}}\right) \\
=\sum_{i=1}^{l}\frac{1}{l}\left(  \hat{p}\left(  i\right)  -\frac{1}%
{l}\right)  f^{\prime}\left(  \frac{\alpha\hat{p}\left(  i\right)  +\left(
1-\alpha\right)  \frac{1}{l}}{\frac{1}{l}}\right)
\end{multline*}
and%
\begin{multline*}
\frac{d}{d\alpha}D\left(  \alpha Emp_{n}\left(  \omega\right)  +\left(
1-\alpha\right)  U\Vert U\right) \\
=\sum_{i=1}^{l}\frac{1}{l}\left(  \hat{p}\left(  i\right)  -\frac{1}%
{l}\right)  ^{2}f^{\prime\prime}\left(  \frac{\alpha\hat{p}\left(  i\right)
+\left(  1-\alpha\right)  \frac{1}{l}}{\frac{1}{l}}\right)  .
\end{multline*}
Thus a second order Taylor expansion gives%
\begin{align*}
D\left(  \alpha Emp_{n}\left(  \omega\right)  +\left(  1-\alpha\right)  U\Vert
U\right)   &  \approx\frac{f^{\prime\prime}\left(  1\right)  }{2}\sum
_{i=1}^{l}\frac{1}{l}\left(  \hat{p}\left(  i\right)  -\frac{1}{l}\right)
^{2}\\
&  =\frac{\chi^{2}\left(  Emp_{n}\left(  \omega\right)  ,U\right)  }{2l^{2}}.
\end{align*}
Thus using a small value of $\alpha$ approximately corresponds to replace the
likelihood ratio test with a $\chi^{2}$ test.
\end{example}

\begin{example}
[Normality test continued]Small or large values of $\beta$ corresponds to
small or large values of $\alpha.$ If the $i^{\prime}$th observation is
denoted $x_{i}$ then
\begin{multline*}
D\left(  \Psi_{d_{0}}\left(  Emp_{n}\left(  \omega\right)  \right)  \Vert
\Psi_{d_{0}}\left(  \Phi\right)  \right) \\
=D\left(  \Psi_{d_{0}}\left(  \frac{1}{n}\sum_{i=1}^{n}\delta_{x_{i}}\right)
\Vert\Psi_{d_{0}}\left(  \Phi\right)  \right) \\
=D\left(  \frac{1}{n}\sum_{i=1}^{n}\Psi_{d_{0}}\left(  \delta_{x_{i}}\right)
\Vert\Psi_{d_{0}}\left(  \Phi\right)  \right) \\
=\frac{1}{n}\sum_{i=1}^{n}D\left(  \Psi_{d_{0}}\left(  \delta_{x_{i}}\right)
\Vert\Psi_{d_{0}}\left(  \Phi\right)  \right) \\
+\frac{1}{n}\sum_{i=1}^{n}D\left(  \Psi_{d_{0}}\left(  \delta_{x_{i}}\right)
\Vert\frac{1}{n}\sum_{j=1}^{n}\Psi_{d_{0}}\left(  \delta_{x_{j}}\right)
\right) \\
=\frac{1}{n}\sum_{i=1}^{n}\frac{x_{i}^{2}}{2}\\
+\frac{1}{n}\sum_{i=1}^{n}D\left(  \Psi_{d_{0}}\left(  \delta_{x_{i}}\right)
\Vert\frac{1}{n}\sum_{j=1}^{n}\Psi_{d_{0}}\left(  \delta_{x_{j}}\right)
\right)  .
\end{multline*}
For large values of $\alpha$ we only smooth a little so the different
observations smoothed are approximately singular. Thus%
\[
D\left(  \Psi_{d_{0}}\left(  Emp_{n}\left(  \omega\right)  \right)  \Vert
\Psi_{d_{0}}\left(  \Phi\right)  \right)  \approx\frac{\frac{1}{n}\sum
_{i=1}^{n}x_{i}^{2}}{2}+\log n.
\]
In this case the use of rate distortion statistic is approximately equivalent
to the use of the statistic $\frac{1}{n}\sum_{i=1}^{n}x_{i}^{2}.$ This
statistic is sufficient for alternatives in the exponential family
$\Phi\left(  0,\sigma^{2}\right)  ,\sigma>0.$

For small values of $\alpha$ we use a different expansion. We use that
$D\left(  \alpha X+\left(  1-\alpha^{2}\right)  ^{1/2}Z\Vert\Phi\right)  $ has
a leading term determined by the mean value of $X.$ Therefore the statistic
essentially reduces to $\frac{1}{n}\sum_{i=1}^{n}x_{i}^{2}.$ This statistic is
sufficient for alternatives in the exponential family $\Phi\left(
\mu,1\right)  .$
\end{example}

\begin{example}
[Uniformity of angular data continued]Small or large values of $\beta$
corresponds to small or large values of $\kappa.$ For small values of $\kappa$
we have
\[
\frac{\exp\left(  \kappa\cos\left(  \theta\right)  \right)  }{2\pi
I_{0}\left(  \kappa\right)  }\approx1+\kappa\cos\left(  \theta\right)  .
\]
For observations $\theta_{1},\theta_{2},...,\theta_{n}$ the smoothed
distribution approximately has density
\begin{multline*}
\frac{1}{n}\sum_{i=1}^{n}\left(  1+\kappa\cos\left(  \theta-\theta_{1}\right)
\right) \\
=1+\frac{\kappa}{n}\sum_{i=1}^{n}\left(  \cos\left(  \theta-\theta_{i}\right)
\right) \\
=1+\frac{\kappa}{n}\sum_{i=1}^{n}\left(
\begin{array}
[c]{c}%
\cos\theta\\
\sin\theta
\end{array}
\right)  \cdot\left(
\begin{array}
[c]{c}%
\cos\theta_{i}\\
\sin\theta_{i}%
\end{array}
\right) \\
=1+\kappa\left(
\begin{array}
[c]{c}%
\cos\theta\\
\sin\theta
\end{array}
\right)  \cdot\frac{1}{n}\sum_{i=1}^{n}\left(
\begin{array}
[c]{c}%
\cos\theta_{i}\\
\sin\theta_{i}%
\end{array}
\right)  .
\end{multline*}
The rate distortion statistic will approximately be given by $\frac{1}{n}%
\sum_{i=1}^{n}\left(
\begin{array}
[c]{c}%
\cos\theta_{i}\\
\sin\theta_{i}%
\end{array}
\right)  .$ By rotational symmetry the information divergence does not depend
on the direction of the vector $\frac{1}{n}\sum_{i=1}^{n}\left(
\begin{array}
[c]{c}%
\cos\theta_{i}\\
\sin\theta_{i}%
\end{array}
\right)  .$ Thus the use of the rate distortion statistic is essentially
equivalent to the use of the statistic $\left\Vert \frac{1}{n}\sum_{i=1}%
^{n}\left(
\begin{array}
[c]{c}%
\cos\theta_{i}\\
\sin\theta_{i}%
\end{array}
\right)  \right\Vert _{2}^{2}.$ This is the most used statistic for testing
uniformity of angular data.

We have
\begin{multline*}
D\left(  \Psi_{d_{0}}\left(  Emp_{n}\left(  \omega\right)  \right)  \Vert
U\right) \\
=\frac{1}{n}\sum_{i=1}^{n}D\left(  \Psi_{d_{0}}\left(  \delta_{\theta_{i}%
}\right)  \Vert U\right) \\
+\frac{1}{n}\sum_{i=1}^{n}D\left(  \Psi_{d_{0}}\left(  \delta_{\theta_{i}%
}\right)  \Vert\frac{1}{n}\sum_{j=1}^{n}\Psi_{d_{0}}\left(  \delta_{x_{j}%
}\right)  \right) \\
=D\left(  \Psi_{d_{0}}\left(  \delta_{0}\right)  \Vert U\right) \\
+\frac{1}{n}\sum_{i=1}^{n}D\left(  \Psi_{d_{0}}\left(  \delta_{\theta_{i}%
}\right)  \Vert\frac{1}{n}\sum_{j=1}^{n}\Psi_{d_{0}}\left(  \delta_{x_{j}%
}\right)  \right)  .
\end{multline*}
For large values of $\kappa$ the term
\[
\frac{1}{n}\sum_{i=1}^{n}D\left(  \Psi_{d_{0}}\left(  \delta_{\theta_{i}%
}\right)  \Vert\frac{1}{n}\sum_{j=1}^{n}\Psi_{d_{0}}\left(  \delta_{x_{j}%
}\right)  \right)
\]
will be dominated by the pair $\left(  \theta_{i},\theta_{j}\right)  ,i\neq j$
for which $\cos\left(  \theta_{i}-\theta_{j}\right)  $ is maximal.
\end{example}

\section{Hodge and Lehman efficiency}

For testing uniformity with Hamming distortion we see that if we do not
compress data ($\alpha=1$) the rate distortion test gives the statistic
$D\left(  Emp_{n}\left(  \omega\right)  \Vert U\right)  $ which is known to be
Bahadur efficient for testing uniformity. This is in general not the case. For
a rate distortion test of normality little compression gives a statistic that
is efficient for Gaussian alternatives with mean zero and variance different
from 1, but it is obviously not efficient against other alternatives with mean
0 and variance 1. Similarly the rate distortion test of uniformity of angular
data depends of the maximal value of $\cos\left(  \theta_{i}-\theta
_{j}\right)  ,i\neq j$ but not on values of all other observed angles which is
obviously not efficient. So the question is how much one should to compress in
order to get an efficient test against any alternative.

There are several ways of measuring efficiency among which the following are
most important. In this short note it is neither possible to give all
definitions nor proofs in details.

\textbf{Hodge and Lehman efficiency} An alternative hypothesis and a
significance level are fixed. One is interested in the sample size that is
needed to achieve a certain large power of the test.

\textbf{Bahadur efficiency} An alternative hypothesis and a power level are
fixed. One is interested in the sample size that is needed to achieve a
certain small significance level of the test.

\textbf{Pitman efficiency} The alternative is moved closer when the sample
size is increased. This is done in a way so that the power of the test is
constant. One is interested in the sample size that is needed to achieve a
certain fixed significance level of the test.

The Hodge and Lehman efficiency is often the easiest to calculate but most
tests are equally efficient in this sense. More tests can be distinguished by
their Pitman efficiency. The Bahadur efficiency is often the most sensitive
and at the same time often the hardest to calculate.

\begin{theorem}
Assume that the space $\Omega$ is compact and that the distortion function is
continuous. Let $d_{n}$ denote a decreasing sequence of distortion values.
Assume that $Q$ generates data. Then
\[
Q\left(  D\left(  \Psi_{d_{n}}\left(  Emp_{n}\left(  \omega\right)  \right)
\Vert\Psi_{d_{n}}\left(  Q\right)  \right)  \geq\varepsilon\right)
\rightarrow0\text{ for }n\rightarrow\infty
\]
if $d_{n}$ tends to $0$ sufficiently slowly.
\end{theorem}

\begin{proof}
It is sufficient to show that
\[
Q\left(  D\left(  \Psi_{\delta}\left(  Emp_{n}\left(  \omega\right)  \right)
\Vert\Psi_{\delta}\left(  Q\right)  \right)  \geq\varepsilon\right)
\rightarrow0
\]
for any fixed distortion level $\delta>0.$ Weak convergence means that
$Emp_{n}\left(  \omega\right)  $ converges to $Q$ in the Wasserstein sense.
Continuity of the distortion function $d$ implies that $\Psi_{\delta}$ is weak
continuous on the set of probability measures.
\end{proof}

\begin{theorem}
\label{Sanov}Let $d_{n}$ denote a decreasing sequence of distortion values.
Assume that $Q$ generates data. Then
\[
\lim\inf D\left(  \Psi_{d_{n}}\left(  Emp_{n}\left(  \omega\right)  \right)
\Vert\Psi_{d_{n}}\left(  P\right)  \right)  \geq D\left(  Q\Vert P\right)
\]
almost surely.
\end{theorem}

\begin{proof}
This follows by lower semi-continuity of information because $\Psi_{d_{n}%
}\left(  Emp_{n}\left(  \omega\right)  \right)  $ tends to $Q$ and
$\Psi_{d_{n}}\left(  P\right)  $ tends to $P$ in the weak topology.
\end{proof}

If $P$ denotes an alternative to a nul-hypothesis $Q$ then according to
Sanov's theorem for a fixed significance level the best achievable type 2
error decreases like $\exp\left(  -nD\left(  Q\Vert P\right)  \right)  .$ The
two previous theorems together implies that the rate-distortion test on a
compact set with a continuous distortion function achieves the same
exponential decrease in type 2 error. Hence, the rate distortion test is
efficient in the sense of Hodge and Lehman.

\section{Bahadur efficiency}

We shall analyze this question in the case of testing uniformity of angular
data because this is of particular simplicity because angles can be identified
with elements of $SO\left(  2\right)  $.

\begin{theorem}
Let $d_{n}$ denote a decreasing sequence of distortion values. Assume that $P$
generates data. Then
\[
\lim\inf D\left(  \Psi_{d_{n}}\left(  Emp_{n}\left(  \omega\right)  \right)
\Vert U\right)  \geq D\left(  P\Vert U\right)
\]
almost surely.
\end{theorem}

\begin{proof}
The proof is essentially the same as the proof of Theorem \ref{Sanov}.
\end{proof}

The theorem implies that for any $K<D\left(  P\Vert U\right)  $ we have
$D\left(  \Psi_{d_{n}}\left(  Emp_{n}\left(  \omega\right)  \right)  \Vert
U\right)  \geq K$ eventually almost surely so if $P$ is the distribution of
the alternative hypothesis then and the power of the test is kept fixed, then
the acceptance regions of alternative $P$ in the rate distortion test must
have the form $D\left(  \Psi_{d_{n}}\left(  Emp_{n}\left(  \omega\right)
\right)  \Vert U\right)  \geq K_{n}$ for $K_{n}\rightarrow D\left(  P\Vert
U\right)  .$ In order to determine the Bahadur efficiency we have to bound the
probability of $D\left(  \Psi_{d_{n}}\left(  Emp_{n}\left(  \omega\right)
\right)  \Vert U\right)  \geq K_{n}$ under the null hypothesis that data are
generated by a uniform distribution. Now partition the set of angles $\left[
0;2\pi\right[  $ into $k_{n}$ intervals of length $2\pi/k_{n}.$ We choose
$k_{n}$ such that $\frac{n}{k\log k}\rightarrow\infty$ for $n\rightarrow
\infty.$ Let $F_{n}$ denote the $\sigma$-algebra generated by these intervals.
Then
\[
\lim-\frac{1}{n}\Pr\left(  D\left(  Emp_{n}\left(  \omega\right)  _{\mid
F_{n}}\Vert U_{\mid F_{n}}\right)  \geq K_{n}\right)  =D\left(  P\Vert
U\right)  .
\]
We are interested in
\[
D\left(  \Psi_{d_{n}}\left(  Emp_{n}\left(  \omega\right)  \right)  \Vert
U\right)  =D\left(  \Psi_{d_{n}}\left(  Emp_{n}\left(  \omega\right)  \right)
\Vert\Psi_{d_{n}}\left(  U\right)  \right)
\]
and not $D\left(  Emp_{n}\left(  \omega\right)  _{\mid F_{n}}\Vert U_{\mid
F_{n}}\right)  $ but each subinterval has length $2\pi/k_{n}$ so
\begin{multline*}
\left\vert \log\frac{d\Psi_{d_{n}}\left(  Emp_{n}\left(  \omega\right)
\right)  }{d\Psi_{d_{n}}\left(  U\right)  }-\log\frac{dEmp_{n}\left(
\omega\right)  _{\mid F_{n}}}{dU_{\mid F_{n}}}\right\vert \\
\leq\left\vert \log\frac{\frac{\exp\left(  \kappa_{n}\cos\left(  0\right)
\right)  }{2\pi I_{0}\left(  \kappa\right)  }}{\frac{\exp\left(  \kappa
_{n}\cos\left(  \theta\right)  \right)  }{2\pi I_{0}\left(  \kappa\right)  }%
}\right\vert \\
=\kappa_{n}\left\vert \cos\left(  0\right)  -\cos\left(  2\pi/k_{n}\right)
\right\vert .
\end{multline*}
Therefore
\[
\lim-\frac{1}{n}\Pr\left(  D\left(  \Psi_{d_{n}}\left(  Emp_{n}\left(
\omega\right)  \right)  \Vert U\right)  \geq K_{n}\right)  =D\left(  P\Vert
U\right)
\]
if the the test is Bahadur efficient if
\[
\frac{\kappa_{n}\left\vert \cos\left(  0\right)  -\cos\left(  2\pi
/k_{n}\right)  \right\vert }{n}\rightarrow0
\]
for $n\rightarrow\infty.$ An expansion of cosine around $0$ shows that the
condition is equivalent to
\[
\frac{\kappa_{n}}{nk_{n}^{2}}\rightarrow0\text{ for }n\rightarrow\infty.
\]
If we choose $k_{n}=n^{\gamma}$ where $\gamma<1$ we get the sufficient
condition%
\[
\frac{\kappa_{n}}{n^{1+2\gamma}}\rightarrow0\text{ for }n\rightarrow\infty.
\]
This leads us to the following theorem.

\begin{theorem}
The rate distortion test of uniformity of angular data has smoothing by a von
Mises distribution with parameter $\kappa_{n}.$ If $\kappa_{n}\rightarrow
\infty$ for $n\rightarrow\infty$ and there exist $\eta\in\left[  1;3\right[  $
such that
\[
\frac{\kappa_{n}}{n^{\eta}}\rightarrow0\text{ for }n\rightarrow\infty,
\]
then the rate distortion test is Bahadur efficient.
\end{theorem}

The method sketched here can be extended to prove Bahadur efficiency of rate
distortion test of the uniformity on compact groups \cite{Harremoes2006,
Harremoes2009b}.

\section{Discussion}

A new statistical test is proposed. It is based on a rate distortion function.
By specifying the distortion function one does not have to divide the data
into bins as this is build into the test. We have discussed the test in detail
for a few examples. The example with testing uniformity of angular data can be
extended to compact groups. There is no standard procedure for testing
uniformity on a group, but there are many competing tests for the Gaussian
distribution. In \cite{Stephens1974}, \cite{Arizono1989} and
\cite{Steinberg1992} it has been shown by simulations that tests based on
estimation entropy are more powerful than many other test for normality that
one can find in the literature. The author has done some simulation to compare
these tests with the test proposed here. These simulations indicates that the
rate distortion test has a good power, but these results are still preliminary
and will not be presented in this short note.

We saw that the rate distortion test has good Bahadur efficiency for angular
data. We conjecture that the proposed test has high Bahadur efficiency in any
case where it can be applied. It is not clear how to formulate this conjecture
precisely, and it may be hard to prove because the rate distorting function
normally cannot be calculated exactly.

A nice feature about the rate distortion test is that one can get a clear
understanding of the effect of very small or very large compression. In our
examples very small or very large compression in the rate distortion test
corresponds to other familiar test like $\chi^{2}$-testing, and this may
actually be used to give new interpretations of these tests. This is in
contrast with the common approach via discretizations. It is simply difficult
to analyze the effect of discretize data into very few bins because 2 gives an
absolute lower bound on how many bins one can use if the analysis should not
become trivial.

Another conjecture that has been supported by numerical calculations is that
the rate distortion statistics is asymptotically Gaussian. As it is now we
have to Monte Carlo simulate the rate distortion statistics, and each
simulation involves a numerical calculation of the rate distortion function.
If it can be proved that the distribution of the rate distortion statistics is
asymptotically Gaussian it means that the number of simulations can be reduces
significantly because one just has to estimate mean and variance in order to
be able to calculate the critical value for a specified significance level.

In this paper some simple examples where the rate distortion function can be
calculated exactly, have been discussed. There are other examples than these
where the rate distortion function can be calculated exactly. One interesting
example is the Poisson process discussed in \cite{Verdu1996}. The setup is
slightly different than the one presented here and therefore we cannot discuss
it in this short paper. Nevertheless the ideas presented in this paper can be
used to construct a test of whether a random process is a Poisson process.
Contrary to the examples discussed in this paper this test of the Poisson
process is completely new in the sense that it does not relate to any
established statistical test.

\textbf{Acknowledgement} The authors want to thank Peter Gr\"{u}nwald, Igor
Vajda, and Nisheeth Srivastava for useful discussions.\newpage

\bibliographystyle{ieeetr}
\bibliography{database}

\end{document}